\documentclass[aps,prd,showpacs,amsmath,amssymb,nofootinbib,twocolumn]{revtex4-1}
\usepackage{amsmath,amssymb,graphics,epsfig,subfigure,simplewick}
\usepackage{graphicx}
\usepackage{graphicx,array,dcolumn}
\usepackage{calc,tabularx, epsfig,mathrsfs}
\usepackage{hyperref}
\usepackage{natbib}

\usepackage{amsmath,verbatim,enumerate}
\usepackage{amssymb}
\usepackage{multirow}

\usepackage{bm}

\begin{document}
\title{Quantized \textbf{Scalar} Fields in Curved Spacetime Background of Thick Brane }
\author{Jian Wang\footnote{wangjian16@lzu.edu.cn}}
\author{Yu-Xiao Liu\footnote{liuyx@lzu.edu.cn, corresponding author}}

\address{Institute of Theoretical Physics \& Research Center of Gravitation, Lanzhou University, Lanzhou 730000, China\\
Key Laboratory for Magnetism and Magnetic Materials of the Ministry of Education, Lanzhou University, Lanzhou 730000, China}

\begin{abstract}
In this paper, we adopt the method of quantum fields in curved spacetime to quantize a free scalar matter field in the braneworld background whose warped factor is of the form that could generate P\"{o}schl-Teller potential. Then we consider the interaction between the scalar field $\phi(x)$ and a classical scalar source $\rho(x)$ with the form $\mathcal{H}_I=\sqrt{-g}\tilde{g}\rho(x)\phi(x)$. The corresponding S-matrix is given and the number of particles generated by the source is obtained. Furthermore, we get the {particle numbers} in the cases that only ground state mode of the field could be detected and both the ground and first exited states could be detected, respectively. Finally, by the particle number density we define, we show how the extra dimension makes a difference specifically by the excited modes.
\end{abstract}

\maketitle

\section{Introduction}
One of the most well-known extra dimension theories,
the Kaluza-Klein(KK) theory \cite{Kaluza:1921tu,Klein:1926fj}, was proposed as an extension
of Einstein's general relativity to unify {four-dimensional} gravity and electromagnetism. In the KK theory, the extra dimension is compact and the matter fields can propagate in the extra dimension.
{While in Arkani-Hamed-Dimopoulos-Dvali (ADD) brane world} model \cite{ArkaniHamed:1998rs}, matter fields are supposed to be {localized} on a sub-manifold (called brane) {embedded in} the bulk while gravity
could propagate in the bulk. In the ADD model the four-dimensional Planck scale is the product of the volume of {the compact} extra dimensions and the fundamental scale of gravity. This implies that the characteristic quantum-gravity scale could be as low as the weak scale \cite{Giudice:1998ck}. By introducing the {KK} decomposition, the {KK} states of {the graviton can be obtained and they will} give contribution to
the coupling constants in the interaction. Furthermore, the creation and
annihilation of {the KK gravitons} and the interaction between the KK {gravitons} and matter fields {were investigated} \cite{Giudice:1998ck, Han1999, Mirabelli:1998rt, Hewett:1998sn,Cheung:1999zw}. {Different from the ADD model}, the extra dimension in {the Randall-Sundrum (RS)} model
could be infinite \cite{Randall:1999ee,Randall:1999vf}. Besides, the gauge
hierarchy problem could be solved by the
effect of the warped extra dimension in a more satisfactory way. The interaction between gravitons and matter fields in the RS model {was also investigated in Refs.} \cite{Davoudiasl:1999jd, KumarRai:2003kk}.

Based on the
RS model and domain wall model \cite{Rubakov:1983bb}, thick brane models
{were introduced later. In} the thick brane model, matter
fields are not necessarily on the brane \cite{DeWolfe:1999cp,Dvali:2000hr,Gremm:2000dj}.
{Many works have focused on various aspects of thick brane models, see the recent review paper \cite{Liu:2017gcn} and references therein for more details. In a thick brane model, just like gravity, any matter field is higher-dimensional and spreads in the bulk. The zero modes of the matter fields localized on the branes denote the four-dimensional fields in the standard model (SM), while the massive KK modes are new particles beyond the SM.}
{However, as far as we know, few work has focused on the interaction between the KK modes of matter fields or between the KK gravitons and matter fields at the quantum field theory level in thick brane models.}
In the thick brane, if we write the metric as the perturbation with a
Minkowski metric, we get a five-dimensional free matter field  and the coupling between this free field and the perturbation. Further, if we employ the perturbation procedure to solve this interaction, we need to quantize this five-dimensional free matter field and we will have difficulty in explaining the localization of this free matter field on the brane. Also the five-dimensional free matter field has a different mass-energy relation from the usual four-dimensional one. These inspire us to find an alternative method to investigate the interaction between matter fields in the presence of gravity or that between gravity and matter fields in thick brane models at the quantum level.

On the other hand, quantum field theory in curved
spacetime has succeeded in dealing with the quantum
phenomena in the background of curved spacetime. For example, besides the famous Unruh effect \cite{Unruh:1976db} and Hawking radiation\cite{Hawking1975}, the method of quantum fields in curved spacetime {was} widely used in {the problem such as} the particle pair production in the de Sitter spacetime due to an electric field \cite{Garriga:1994bm,Villalba:1995za}. As for the extra dimension theory, the $\lambda\phi^4$ theory in compact {$D$-dimensional} spacetime \cite{Fucci:2017weg} and the effective action of scalar matter fields in the thick brane model \cite{daSilva:2015jpa} {were} investigated in the way of quantum fields in curved spacetime. Also, quantum field theory in curved spacetime has been well developed itself as an axiomatic theory by the algebraic quantum field way in recent decades \cite{Hollands:2014eia}. This indicates that we can adopt quantum field theory in curved spacetime to consider the quantization of matter fields in the thick brane scenario. Moreover, with the quantization of fields, we can use the perturbation procedure to investigate the interaction of matter fields in curved spacetime. In this method {gravity} participates in the interaction as a curved background that affects the interaction process rather than the usual gravitons. We should note that quantum field theory in curved spacetime treats {gravity} classically, i.e., {gravity} is not quantized. Therefore, it is not an ultimate theory of quantum gravity \cite{Hollands:2014eia} and this imposes some limits (for example, the effects of quantum gravity can be neglected) on our thick brane model under which this method is available. Although the method we adopt is half classical ({gravity} is not quantized), it is not meaningless. On the contrary, because of the absence of a well convincing quantum gravity theory, we could use quantum field theory in curved spacetime to give some results that could be treated as a classical limit that the potential quantum gravity should give. The results may be also criteria for the potential quantum gravity.

In this paper, we consider the quantization of scalar matter fields in the thick brane scenario where the background spacetime is of the type that could generate P\"{o}schl-Teller potential. The interaction linear in the scalar fields is investigated later. Furthermore, by the perturbation procedure, we could solve this interaction and then give a hint to detect extra dimensions with the excited states of matter fields.

This paper is organized as follows. In Sec \ref{secII}, we consider the solution of scalar matter fields in the thick brane scenario. We take the background spacetime metric to be the form that could generate the P\"{o}schl-Teller potential. Next, we give the quantization of scalar fields by the way of quantum fields in curved spacetime. Then in Sec \ref{secIII}, based on our quantized fields, we get the commutation relation of the fields and the Feynman propagator in the case of five-dimensional spacetime. {We consider the self-interaction $\mathcal{H}_I=\sqrt{-g}\tilde{g}\rho(x)\phi(x)$ and give} the analytic form of $S$ matrix. The expectation values of the particle numbers of the ground and first excited {states} of fields are subsequently obtained, with a guide to detect the extra dimension by the excited states in the end. In sec \ref{secIV}, a brief summary and an outlook are given.

\section{Quantization of scalar fields in thick brane scenario}\label{secII}

In a thick brane model, the brane can be generated by various background fields. Thus, various types of braneworld can be constructed. The metric of the braneworld scenario can be usually written as
\begin{equation}
ds^2=e^{2A(y)}\eta_{\mu\nu}dx^\mu dx^\nu-dy^2,
\end{equation}
where $\eta_{\mu\nu}=\text{diag}(1,-1,-1,-1)$ and the warped factor $A(y)$ is a function of the extra dimension $y$.
In a thick brane scenario, various of matter fields propagate in the bulk and their zero modes correspond to particles in the standard model.

In this background, we consider a massive scalar field with the Lagrangian
\begin{equation}
\mathcal{L}=\sqrt{-g}\left(\frac{1}{2}g^{\mu\nu}\nabla_\mu\phi\nabla_\nu\phi-\frac{1}{2}m^2\phi^2 \right).
\end{equation}
Then by varying with respect to $\phi$ and considering the form of metric, we can get the equation of motion
\begin{equation}
\partial^2_t \phi- \partial_i \partial_i \phi=e^{-2A}\partial_y(e^{4A}\partial_y\phi)-e^{2A}m^2\phi,
\end{equation}
where $i$ runs from 1 to 3.
We introduce the decomposition $\phi=\xi(x^\mu)\varphi(y)$. The equations become
\begin{eqnarray}
\partial^2_t \xi- \partial_i \partial_i \xi &=& -\mu^2\xi,\\
e^{-2A}\partial_y(e^{4A}\partial_y\varphi)-e^{2A}m^2\varphi+\mu^2\varphi &=& 0. \label{eq4}
\end{eqnarray}
After the coordinate transformation $dz=e^{-A(y)}dy$, Eq. \eqref{eq4} becomes
\begin{equation}
\frac{d^2\varphi}{dz^2}+3e^{2A}\frac{dA}{dz}\frac{d\varphi}{dz}-e^{2A}m^2\varphi +\mu^2\phi=0. \label{eq5}
\end{equation}
By making the field transformation $\varphi=\psi(z)\gamma(A)$ with $\gamma=e^{-\frac{3}{2}A}$, Eq. \eqref{eq5} becomes
\begin{equation}\label{eqofex}
-\frac{d^2\psi}{dz^2}
 +\left(\frac{9}{4}\left(\frac{dA}{dz}\right)^2+\frac{3}{2}\frac{d^2A}{dz^2} +e^{2A}m^2\right)\psi=\mu^2\psi.
\end{equation}

Now we will proceed with a specific braneworld scenario, i.e., a specific form of warped factor and investigate the quantization of a free scalar field in this background spacetime. As for how to construct the braneworld, we neglect the procedure and it can be seen in Ref. \cite{Liu:2017gcn, Gu:2018lub, Gu:2016nyo, Cui:2018hqx}.
In this paper, we consider the following warped factor
\begin{equation}
A(z)=\ln(b\,\text{sech}(a z)),
\end{equation}
for which the spacetime is asymptotically Anti-de Sitter.
Then we can get the Schr\"{o}dinger-like equation with a P\"{o}schl-Teller potential from Eq. (\ref{eqofex})
\begin{equation}
 \left(- \frac{1}{a^2}\frac{d^2}{d{z}^2}
   + V_{\texttt{PT}} \right) \psi(z)=\frac{\mu^2}{a^2} \psi(z), \label{SChrodingEq}
\end{equation}
 where the effective potential is given by
\begin{equation}
 V_{\texttt{PT}}= \frac{9}{4}
        +\left(\frac{b^2m^2}{a^2}-\frac{15}{4}\right)\text{sech}^2(az).  \label{VPT}
\end{equation}
The general solution of Eq.~\eqref{SChrodingEq} is \cite{Guo:2011qt}
\begin{eqnarray}
\psi(z)&=& d_1 P^{u}_{v}\big(\tanh(a z)\big)
    +d_2 Q^{u}_{v}\big(\tanh(a z)\big).
\end{eqnarray}
where $P^{u}_{v}(z)$ and $Q^{u}_{v}(z)$ are  the associated Legendre functions of the first and second kinds, respectively, $d_1$ and $d_2$ are arbitrary constants  and
\begin{eqnarray}
{u}&=&\sqrt{\frac{9}{4}-\frac{\mu^2}{a^2}}, \\
{v} &=& \sqrt{4-\frac{b^2m^2}{a^2}}  -\frac{1}{2}.
\end{eqnarray}
The above solution can be also written as
\begin{equation}\label{generalsol}
\psi(z)= d'_1 P^{u}_{v}\big(\tanh(a z)\big)
    +d'_2 P^{-u}_{v}\big(\tanh(a z)\big),
\end{equation}
with
\begin{eqnarray}
Q^{u}_{v}(x) &=& \frac{\pi}{2\sin(u \pi)} \bigg[ P^u_v(x)\cos( u\pi) \nonumber \\
    &&-\frac{\Gamma(v+u+1)}{\Gamma(v-u+1)}P^{-u}_v(x) \bigg],
\end{eqnarray}
where $\Gamma(u)$ is the Euler gamma function. Considering that $P^{-u}_{v}$ and $P^{u}_{v}$ have good asymptotic forms, we will use the form of solution (\ref{generalsol}) in the rest of this paper.

Next we will go to a simple case of a massless scalar field, i.e., $m=0$, for which the effective potential \eqref{VPT} is simplified as
\begin{equation}
 V_{\texttt{PT}}= \frac{9}{4}  -\frac{15}{4} \text{sech}^2(a z).  \label{VPTmassless}
\end{equation}
The solutions contain two bound states
\begin{eqnarray}
\psi_0 &\propto& \text{sech}^{\frac{3}{2}}(a z),~~~~~~~~~~~~~~~(\mu =0)\\
\psi_1 &\propto& \sinh(a z)\text{sech}^{\frac{3}{2}}(a z),~~~~(\mu =\sqrt{2}a)
\end{eqnarray}
and a series of continuous states with $\mu$ beginning from $\frac{3 a}{2}$
\begin{eqnarray}
\psi(z) =  C_1 P^{i\lambda}_{{3}/{2}}\big(\tanh(az)\big)
   +C_2 P^{-i\lambda}_{{3}/{2}}\big(\tanh(az)\big),
\end{eqnarray}
where $C_1$ and $C_2$ are arbitrary constants and
\begin{eqnarray}
   \lambda=\sqrt{\frac{\mu^2}{a^2}-\frac{9}{4}}.
\end{eqnarray}

With $\gamma(z)=e^{-\frac{3}{2}A}=b^{-\frac{3}{2}}\text{sech}^{-\frac{3}{2}}(az)$,
we can write the two bound states as
\begin{eqnarray}
\varphi_0 &=&  c_1,\label{classicalsolution1}\\
\varphi_1 &=& c_2 \sinh(a z).\label{classicalsolution2}
\end{eqnarray}
The continuous states can be also given as
\begin{eqnarray}\label{classicalsolution3}
\varphi&=&\text{sech}^{-\frac{3}{2}}(az)\bigg(C'_{1} P^{i\lambda}_{{3}/{2}}\big(\tanh(az)\big)\nonumber\\
    &&+C'_{2} P^{-i\lambda}_{{3}/{2}}\big(\tanh(az)\big)\bigg).
\end{eqnarray}
In addition, $c_1$, $c_2$, $C'_1$ and $C'_2$ are arbitrary constants.

Then we can go to the quantization of fields. For the scenario we consider in this paper, the spacetime is static and in this case the concept of particles is well defined \cite{Birrell:1982ix,Crispino2008}. Thus we can follow the quantization procedure of scalar fields in curved spacetime. One can consult Refs. \cite{Birrell:1982ix} and \cite{Crispino2008} for details.  First, for the solutions (\ref{classicalsolution1}), (\ref{classicalsolution2}) and (\ref{classicalsolution3}) to the Klein-Gordon equation, we need to determine the constants $c_1$, $c_2$, $C'_1$ and $C'_2$ so that the K-G inner product of the solutions could satisfy the conditions
\begin{eqnarray}
(\phi_i,\phi_j)_{\text{KG}}=-(\phi^*_i,\phi^*_j)_{\text{KG}}=\delta_{ij},\label{requirements1}\\
(\phi^*_i,\phi_j)_{\text{KG}}=(\phi_i,\phi^*_j)_{\text{KG}}=0,\label{requirements2}
\end{eqnarray}
where the K-G inner product is defined as
\begin{equation}
(\phi_A,\phi_B)_{\text{KG}}=i\int d^3x dy e^{2A}(\phi^*_A\partial_t \phi_B-\phi_B\partial_t \phi^*_A),
\end{equation}
in our scenario. This implies that $\varphi(y)$ should obey
\begin{eqnarray}\label{normalization}
\int dy e^{2A} \varphi^*_i(y)\varphi_j(y) &=& \delta_{ij},\\
\int dy e^{2A} \varphi^*_{\mu'}(y)\varphi_\mu(y) &=& \delta(\mu-\mu'),
\end{eqnarray}
for the bound states and continuous states, respectively. Here $i$ and $j$ run from 1 to 2. We should note that for the bound states, the solutions are real and $\varphi^*$ is the complex conjugate of $\varphi$, i,e., $\varphi$ itself. But for the continuous states, the general solutions consist of two independent special solutions. So $\varphi$ and $\varphi^*$ represent these two special solutions, respectively. For Schr\"{o}dinger-like equation (\ref{eq4}), according to Sturm-Liouville theory, the solutions corresponding to different eigenvalues are orthogonal with the weight function $e^{2A}$. This statement is consistent with the conditions (\ref{requirements1}) and (\ref{requirements2}). So we only need to specify the constants $c_1$, $c_2$, $C'_1$ and $C'_2$ by the requirements that $\varphi_\mu$ and $\varphi_i$ are supposed to be normalized with weight function $e^{2A}$, respectively. After calculation, we get that
\begin{eqnarray}
c^2_1=c^2_2=\frac{2a}{\pi b^3},
\end{eqnarray}
and from Ref. \cite{Guo:2011qt} and \cite{virchenko2001generalized}, we could get that
\begin{equation}
\begin{split}
C'_1&=\sqrt{\frac{a}{b^3}}   \frac{\big|\Gamma(1-i\lambda)\big|}{\sqrt{2\pi}}  \\
&=C'_2=\sqrt{\frac{a}{b^3}}\frac{\big|\Gamma(1+i\lambda)\big|}{\sqrt{2\pi}}\\
&=C'.
\end{split}
\end{equation}
Then the quantum field $\phi(x^\mu,z)$ is given as
\begin{equation}
\begin{split}
\phi&=\int\frac{dp^3}{(2\pi)^{\frac{3}{2}}\sqrt{2\omega_{\vec{p},\mu_{(0)}}}}\sqrt{\frac{2a}{\pi b^3}}
   \bigg(e^{i p x} a^\dagger_{\vec{p},\mu_{(0)}}\\
&+e^{-i p x} a_{\vec{p},\mu_{(0)}}\bigg)\\
& + \int\frac{dp^3}{(2\pi)^{\frac{3}{2}}\sqrt{2\omega_{\vec{p},\mu_{(1)}}}}
   \sqrt{\frac{2a}{\pi b^3}}
   \bigg(e^{i p x} a^\dagger_{\vec{p},\mu_{(1)}}\\
&+e^{-i p x} a_{\vec{p},\mu_{(1)}}\bigg) \sinh(a z)\\
&+\int\frac{dp^3}{(2\pi)^{\frac{3}{2}}\sqrt{2\omega_{\vec{p},\mu}}}
  \int d\mu  C' \text{sech}^{-\frac{3}{2}}(a z) \\
&\times\bigg(e^{i p x}P^{i\lambda}_\frac{3}{2}\big(\tanh(a z)\big)a^\dagger_{\vec{p},\mu,-}\\
&+e^{i p x}P^{-i\lambda}_\frac{3}{2}\big(\tanh(a z)\big)a^\dagger_{\vec{p},\mu,+}\\
&+e^{-i p x} P^{-i\lambda}_\frac{3}{2}\big(\tanh(a z)\big)a_{\vec{p},\mu,-}\\
&+e^{-i p x} P^{i\lambda}_\frac{3}{2}\big(\tanh(a z)\big)a_{\vec{p},\mu,+}\bigg).
\end{split}
\end{equation}
The quantum field can propagate in five-dimensional spacetime. Before we touch any realistic physics, we need to explain why the world is detected as a four-dimensional manifold in our scenario.
In the usual thin braneworld scenario, all the matter corresponding to particles in the standard model is confined on the four-dimensional hyper surface. While in the thick brane scenario, the solution of matter fields is also a function of extra dimension. Thus the localization of matter fields which accounts for the observed four-dimensional world is realized by requiring the distribution of matter fields to be concentrated around the origin of extra dimension. But in usual thick brane models, the matter field is not quantized and in the circumstances, it's natural to adopt the distribution of matter fields to illustrate our observed four-dimensional world. As for our special case where the matter fields are quantized, we give a new explanation for the observed four-dimensional spacetime. We can see from the quantum field that it consists of two bound modes and a series of continuous modes. Furthermore, the ground state mode does not vary with the extra dimension. So we can interpret the ground state mode as the particle detected in the four-dimensional experiments. The first excited state and continuous state modes can be regarded as the signal from the extra dimension. With the energy or four-dimensional mass getting higher, we can see that the higher excited modes can be detected, i.e., we have the chance to detect the extra dimension as the energy of the accelerator gets higher. Also, because the ground state mode does not vary with the extra dimension, we get a vanished five-momentum thus the
mass-energy equivalence in four-dimensional spacetime is not broken.

Next we will give the commutation relation between the quantum fields we have obtained above and the corresponding Feymann propagator. Then we will consider the interaction.

\section{Interaction of matter fields in thick brane model}\label{secIII}

The commutation relations between the creation and annihilation operators are given as
\begin{align}
&[a_{\vec{k},\mu}, a^\dagger_{\vec{k'},\mu'}]=\delta^{(3)}(\vec{k}-\vec{k'})\delta_{\mu\mu'},\\
&[a_{\vec{k},\mu}, a_{\vec{k'},\mu'}]=0,
\end{align}
and
\begin{align}
&[a_{\vec{k},\mu,\pm}, a^\dagger_{\vec{k'},\mu',\pm}]=\delta^{(3)}(\vec{k}-\vec{k'})\delta(\mu-\mu'),\\
&[a_{\vec{k},\mu,\pm}, a^\dagger_{\vec{k'},\mu',\mp}]=0,\\
&[a_{\vec{k},\mu,\pm}, a_{\vec{k'},\mu',\pm}]=[a_{\vec{k},\mu,\mp}, a_{\vec{k'},\mu',\mp}]=0,
\end{align}
for the two bound states and continuous states respectively.
Then the commutation relation between the fields could be written as
\begin{equation}\label{cr}
\begin{split}
&\big[\phi(x_1),\phi(x_2)\big]=\int\frac{d^3p_1}{(2\pi)^{\frac{3}{2}}\sqrt{2\omega_{\vec{p}_1,\mu_{(0)}}}}\int\frac{d^3p_2}{(2\pi)^{\frac{3}{2}}\sqrt{2\omega_{\vec{p}_2,\mu_{(0)}}}}\\
&\times\delta^{(3)}(\vec{p}_1-\vec{p}_2)\frac{2a}{\pi b^3}(e^{-ip_1x_1+ip_2x_2}-e^{ip_1x_1-ip_2x_2})\\
&+\int\frac{d^3p_1}{(2\pi)^{\frac{3}{2}}\sqrt{2\omega_{\vec{p}_1,\mu_{(1)}}}}\int\frac{d^3p_2}{(2\pi)^{\frac{3}{2}}\sqrt{2\omega_{\vec{p}_2,\mu_{(1)}}}}\delta^{(3)}(\vec{p}_1-\vec{p}_2)\\
&\times\frac{2a}{\pi b^3}(e^{-ip_1x_1+ip_2x_2}-e^{ip_1x_1-ip_2x_2})\sinh(a z_1)\sinh(a z_2)\\
&+\int\frac{d\mu_1 d^3p_1}{(2\pi)^{\frac{3}{2}}\sqrt{2\omega_{\vec{p}_1,\mu_1}}}\int\frac{d\mu_2 d^3p_2}{(2\pi)^{\frac{3}{2}}\sqrt{2\omega_{\vec{p}_2,\mu_2}}}\delta^{(3)}(\vec{p}_1-\vec{p}_2)\\
&\times\delta(\mu_1-\mu_2)C'^2\text{sech}^{-\frac{3}{2}}(a z_1)\text{sech}^{-\frac{3}{2}}(a z_2)\\
&\times\bigg(e^{-ip_1x_1+ip_2x_2}P^{-i\lambda_1}_{{3}/{2}}\big(\tanh(az_1)\big)P^{i\lambda_2}_{{3}/{2}}\big(\tanh(az_2)\big)\\
&+ e^{-ip_1x_1+ip_2x_2}P^{i\lambda_1}_{{3}/{2}}\big(\tanh(az_1)\big)P^{-i\lambda_2}_{{3}/{2}}\big(\tanh(az_2)\big)\\
&-e^{ip_1x_1-ip_2x_2}P^{i\lambda_1}_{{3}/{2}}\big(\tanh(az_1)\big)P^{-i\lambda_2}_{{3}/{2}}\big(\tanh(az_2)\big)\\
&-e^{ip_1x_1-ip_2x_2}P^{-i\lambda_1}_{{3}/{2}}\big(\tanh(az_1)\big)P^{i\lambda_2}_{{3}/{2}}\big(\tanh(az_2)\big)\bigg).
\end{split}
\end{equation}
Here we have used $\lambda_1$ to denote $\sqrt{\frac{\mu^2_1}{a^2}-\frac{9}{4}}$ and $\lambda_2$ to denote $\sqrt{\frac{\mu^2_2}{a^2}-\frac{9}{4}}$, respectively.

The first term of this expression is similar to the commutation relation of quantum fields in four-dimensional Minkowski spacetime multiplying a coefficient $\frac{2a}{\pi b^3}$. This is in good coincidence with the statement that we explain the ground state as the four-dimensional quantum fields we observe. The rest is contribution of the excited states to the commutation relation. It is easy to see that the extra dimension coordinate only appears in this part. So the effect of extra dimension could be gotten from these terms.

Next, we will give the commutation relation between the generalized momentum and field. For a given Lagrangian, the generalized momentum is
\begin{equation}
\pi^\mu=\frac{\partial \mathcal{L}}{\partial \nabla_\mu\phi}.
\end{equation}
For our free quantum fields, we get
\begin{equation}
\pi^\mu=\sqrt{-g}g^{\mu\nu}\nabla_\nu\phi.
\end{equation}
The zero component gives
\begin{equation}
\pi^0=e^{2A}\partial_t\phi.
\end{equation}
So one can get
\begin{equation}
\big[\pi^0(t,x_1),\phi(t,x_2)\big]=e^{2A}\big[\partial_t\phi(t, x_1),\phi(t,x_2)\big],
\end{equation}
and
\begin{eqnarray}
&&\big[\partial_t\phi(t,x_1),\phi(t,x_2)\big] \nonumber \\
&=& -i \frac{2a}{\pi b^3} \delta^{(3)}(\vec{x}_1-\vec{x}_2) \nonumber\\
&-&i \frac{2a}{\pi b^3} \delta^{(3)}(\vec{x}_1-\vec{x}_2)\sinh(a z_1) \sinh(a z_2) \nonumber\\
&-&i \delta^{(3)}(\vec{x}_1-\vec{x}_2) C'^2\text{sech}^{-\frac{3}{2}}(a z_1)
     \text{sech}^{-\frac{3}{2}}(a z_2) \nonumber\\
&\times&\int d\mu\bigg(P^{i\lambda}_{{3}/{2}}\big(\tanh(az_1)\big)P^{-i\lambda}_{{3}/{2}}\big(\tanh(az_2)\big) \nonumber\\
&+&P^{-i\lambda}_{{3}/{2}}\big(\tanh(az_1)\big)P^{i\lambda}_{{3}/{2}}\big(\tanh(az_2)\big)\bigg).
\end{eqnarray}

From these results, we get the Feynman propagator in our five-dimensional scenario
\begin{eqnarray}
&& \langle 0|T(\phi(x_1)\phi(x_2))|0\rangle\nonumber \\
& =& \frac{2a}{\pi b^3}\int\frac{d^4p_1}{(2\pi)^4}
    \frac{i e^{i p_1 (x_1-x_2)}}
         {p_1^2-\mu_{(0)}^2+i\varepsilon}\nonumber\\
& +&  \frac{2a}{\pi b^3}\int\frac{d^4p_1}{(2\pi)^4}
     \frac{i e^{i p_1 (x_1-x_2)}}
          {p_1^2-\mu_{(1)}^2+i\varepsilon}
      \sinh(a z_1)\sinh(a z_2)\nonumber\\
& +& C'^2\int d\mu\int\frac{d^4p_1}{(2\pi)^4}
     \frac{i e^{i p_1 (x_1-x_2)}}
          {p_1^2-\mu^2+i\varepsilon}\nonumber\\
&&\times \text{sech}^{-\frac{3}{2}}(a z_1)\text{sech}^{-\frac{3}{2}}(a z_2)\nonumber\\
&&\times \bigg(P^{i\lambda}_{{3}/{2}}\big(\tanh(az_1)\big)P^{-i\lambda}_{{3}/{2}}\big(\tanh(az_2)\big) \nonumber\\
&+& P^{-i\lambda}_{{3}/{2}}\big(\tanh(az_1)\big)P^{i\lambda}_{{3}/{2}}\big(\tanh(az_2)\big)\bigg).
\end{eqnarray}
Considering that only the quantization of scalar fields is given above, so here we focus on the type of interaction that only contains scalar fields. Note that any polynomial interaction of degree higher than 4 is not renormalizable. For simplicity, we take the interaction to be $\mathcal{L}_I=-\sqrt{-g}\tilde{g}\rho(x)\phi(x)$, where $\tilde{g}$ is the coupling constant and $\rho(x)$ is a function of spacetime. Also, $\rho(x)$ can be regarded as a source that can create mesons. This can be easily seen from the equation of motion $(\Box+\mu^2)\phi(x)=-\tilde{g}\rho(x)$, comparing with the Maxwell equation with an external source $\Box A^\mu=-e j^\mu$\cite{Coleman:2011xi}. In addition, this type of interaction could be solved exactly.
Note that in perturbation theory Dyson's formula and Wick's theorem are not restricted to the case with specific dimensions. So we can still employ the perturbation procedure to get the solutions to the interaction. For $\mathcal{L}_I=-\sqrt{-g}\tilde{g}\rho(x)\phi(x)$, the Hamiltonian is
\begin{eqnarray}
\mathcal{H}&=&\pi^0\nabla_0\phi-\mathcal{L} \nonumber \\
&=&\frac{1}{2}\sqrt{-g}g^{00}\nabla_0\phi\nabla_0\phi-\frac{1}{2}\sqrt{-g}g^{ij}\nabla_i\phi\nabla_j\phi \nonumber\\
&+&\frac{1}{2}\sqrt{-g}m^2\phi^2+\sqrt{-g}\tilde{g}\rho(x)\phi(x),
\end{eqnarray}
from which we can get the interaction part of the Hamiltonian $\mathcal{H}_I=\sqrt{-g}\tilde{g}\rho(x)\phi(x)$. Then Dyson's formula gives the time evolution operator in the interaction picture $U_I(t,t')=T e^{-i\int^t_{t'} dt''H_I(t'')}$. Expanding $U_I$ by order of $\tilde{g}$, and considering that there is only one contraction of the scalar field, according to Ref. \cite{Coleman:2011xi}, $U_I$ could be given as
\begin{equation}
U_I(\infty,-\infty)=:e^{O_1+\frac{O_2}{2}}:,
\end{equation}
where
\begin{eqnarray}
   O_1 &=& -i \tilde{g}\int d^5 x \sqrt{-g}\rho(x)\phi(x),  \\
   O_2 &=& (-i \tilde{g})^2\int d^5 x_1d^5 x_2 \sqrt{-g(x_1)}\sqrt{-g(x_2)} \nonumber \\
       &\times& \contraction{}{\phi}{(x_1)}{\phi}\phi(x_1)\phi(x_2)\rho(x_1)\rho(x_2)
\end{eqnarray}
with $\contraction{}{\phi}{(x_1)}{\phi}\phi(x_1)\phi(x_2)$  the contraction of the fields.

Using the expression of quantum fields, we get
\begin{equation}
\begin{split}
O_1&=-i\tilde{g}
  \int\frac{dp^3}{(2\pi)^{\frac{3}{2}}\sqrt{2\omega_{\vec{p},\mu_{(0)}}}}
    \Big(\tilde{\rho}(p,\mu_{(0)}) a^\dagger_{\vec{p},\mu_{(0)}}\\
&+\tilde{\rho}(-p,\mu_{(0)}) a_{\vec{p},\mu_{(0)}}\Big)\\
& -i\tilde{g}
  \int\frac{dp^3}{(2\pi)^{\frac{3}{2}}\sqrt{2\omega_{\vec{p},\mu_{(1)}}}}
  \Big(\tilde{\rho}(p,\mu_{(1)})a^\dagger_{\vec{p},\mu_{(1)}}\\
&+\tilde{\rho}(-p,\mu_{(1)})a_{\vec{p},\mu_{(1)}}\Big)\\
& -i\tilde{g} \int\frac{dp^3}{(2\pi)^{\frac{3}{2}}\sqrt{2\omega_{\vec{p},\mu}}}\int d\mu\\
&\times\Big(\tilde{\rho}_+(p,\mu)a^\dagger_{\vec{p},\mu,-}+\tilde{\rho}_-(p,\mu)a^\dagger_{\vec{p},\mu,+}\\
&+\tilde{\rho}_-(-p,\mu)a_{\vec{p},\mu,-}+\tilde{\rho}_+(-p,\mu)a_{\vec{p},\mu,+}\Big) .
\end{split}
\end{equation}
where
\begin{eqnarray}
  \tilde{\rho}(p,\mu_{(0)}) &=& \int dy d^4x \rho(x)e^{4A}e^{ip x}\sqrt{\frac{2a}{\pi b^3}}, \label{pho0}\\
  \tilde{\rho}(p,\mu_{(1)}) &=&  \int dy d^4x \rho(x)e^{4A}e^{ip x}\sqrt{\frac{2a}{\pi b^3}} \sinh(a z),  \label{pho1}\\
  \tilde{\rho}_{\pm}(p,\mu) &=&  \int dy d^4x \rho(x)e^{4A}e^{ip x}C'\text{sech}^{-\frac{3}{2}}(a z) \nonumber \\
  &\times&  P^{{\pm}i\lambda}_\frac{3}{2}\big(\tanh(a z)\big).   \label{phomu}
\end{eqnarray}

On the other hand, considering that the contraction $\contraction{}{\phi}{(x_1)}{\phi}\phi(x_1)\phi(x_2)$ is a c-number, we can take $O_2$ to be $\alpha+i\beta$ with $\alpha$ and $\beta$  real numbers.
Thus, we can get the time evolution operator
\begin{equation}
\begin{split}
&U_I(\infty,-\infty)=:e^{O_1+\frac{O_2}{2}}:\\
&=\exp\bigg[-i\tilde{g}\int\frac{dp^3}{(2\pi)^{\frac{3}{2}}}\bigg(\frac{\tilde{\rho}(p,\mu_{(0)}) a^\dagger_{\vec{p},\mu_{(0)}}}{\sqrt{2\omega_{\vec{p},\mu_{(0)}}}}\\
&+\frac{\tilde{\rho}(p,\mu_{(1)})a^\dagger_{\vec{p},\mu_{(1)}}}{\sqrt{2\omega_{\vec{p},\mu_{(1)}}}}\\
&+\int d\mu\frac{\tilde{\rho}_+(p,\mu)a^\dagger_{\vec{p},\mu,-}
+\tilde{\rho}_-(p,\mu)a^\dagger_{\vec{p},\mu,+}}{\sqrt{2\omega_{\vec{p},\mu}}}\bigg)\bigg]\\
&\times\exp\bigg[-i\tilde{g}\int\frac{dp^3}{(2\pi)^{\frac{3}{2}}}\bigg(\frac{\tilde{\rho}(-p,\mu_{(0)}) a_{\vec{p},\mu_{(0)}}}{\sqrt{2\omega_{\vec{p},\mu_{(0)}}}}\\
&+\frac{\tilde{\rho}(-p,\mu_{(1)})a_{\vec{p},\mu_{(1)}}}{\sqrt{2\omega_{\vec{p},\mu_{(1)}}}}\\
&+\int d\mu\frac{\tilde{\rho}_-(-p,\mu)a_{\vec{p},\mu,-}+\tilde{\rho}_+(-p,\mu)a_{\vec{p},\mu,+}}{\sqrt{2\omega_{\vec{p},\mu}}}\bigg)\bigg]\\
&\times e^{\frac{1}{2}(\alpha+i\beta)}.
\end{split}
\end{equation}
Applying the time revolution operator to the vacuum state, we get
\begin{equation}\label{optovac}
\begin{split}
&U_I(\infty,-\infty)|0\rangle\\
&=e^{\frac{1}{2}(\alpha+i\beta)}\exp\bigg(-i\tilde{g}\int\frac{dp^3}{(2\pi)^{\frac{3}{2}}}\frac{\tilde{\rho}(p,\mu_{(0)}) a^\dagger_{\vec{p},\mu_{(0)}}}{\sqrt{2\omega_{\vec{p},\mu_{(0)}}}}\bigg)\\
&\times\exp\bigg(-i\tilde{g}\int\frac{dp^3}{(2\pi)^{\frac{3}{2}}}\frac{\tilde{\rho}(p,\mu_{(1)}) a^\dagger_{\vec{p},\mu_{(1)}}}{\sqrt{2\omega_{\vec{p},\mu_{(1)}}}}\bigg)\\
&\times\exp\bigg(-i\tilde{g}\int\frac{dp^3 d\mu}{(2\pi)^{\frac{3}{2}}}\frac{\tilde{\rho}_+(p,\mu)a^\dagger_{\vec{p},\mu,-}+\tilde{\rho}_-(p,\mu)a^\dagger_{\vec{p},\mu,+}}{\sqrt{2\omega_{\vec{p},\mu}}}\bigg)|0\rangle\\
&=\sum_{n=0}^\infty\sum_{m=0}^\infty\sum_{l=0}^\infty\sum_{k=0}^\infty
   e^{\frac{1}{2}(\alpha+i\beta)}
   \frac{(-i\tilde{g})^n}{n!}
   \frac{(-i\tilde{g})^m}{m!}
   \frac{(-i\tilde{g})^l}{l!}
   \frac{(-i\tilde{g})^k}{k!}
   \\
&\times  \int\frac{dp_1^3 \tilde{\rho}(p_1,\mu_{(0)})}{(2\pi)^{\frac{3}{2}}\sqrt{2\omega_{\vec{p}_1,\mu_{(0)}}}}\cdot\cdot\cdot\frac{dp_n^3 \tilde{\rho}(p_n,\mu_{(0)})}{(2\pi)^{\frac{3}{2}}\sqrt{2\omega_{\vec{p}_n,\mu_{(0)}}}}\\
&\times \int\frac{dp'^3_1 \tilde{\rho}(p'_1,\mu_{(1)})}{(2\pi)^{\frac{3}{2}}\sqrt{2\omega_{\vec{p'}_1,\mu_{(1)}}}}\cdot\cdot\cdot\frac{dp'^3_m \tilde{\rho}(p'_m,\mu_{(1)})}{(2\pi)^{\frac{3}{2}}\sqrt{2\omega_{\vec{p'}_m,\mu_{(1)}}}}\\
&\times  \int\frac{dp''^3_1 d\mu_1 \tilde{\rho}_+(p''_1,\mu_1)}{(2\pi)^{\frac{3}{2}}\sqrt{2\omega_{\vec{p''}_1,\mu_1}}}\cdot\cdot\cdot\frac{dp''^3_l d\mu_l \tilde{\rho}_+(p''_l,\mu_l)}{(2\pi)^{\frac{3}{2}}\sqrt{2\omega_{\vec{p''}_l,\mu_l}}}\\
&\times \int\frac{dp'''^3_1 d\mu'_1 \tilde{\rho}_-(p'''_1,\mu'_1)}{(2\pi)^{\frac{3}{2}}\sqrt{2\omega_{\vec{p'''}_1,\mu'_1}}}\cdot\cdot\cdot\frac{dp'''^3_k d\mu'_k \tilde{\rho}_-(p'''_k,\mu'_k)}{(2\pi)^{\frac{3}{2}}\sqrt{2\omega_{\vec{p'''}_k,\mu'_k}}}\\
&|p_1(\mu_{(0)}), \cdot \cdot \cdot, p_n(\mu_{(0)}), p'_1(\mu_{(1)}), \cdot\cdot\cdot, p'_m(\mu_{(1)}),\\
&p''_1(\mu_1), \cdot\cdot\cdot, p''_l(\mu_l), p'''_1(\mu'_1), \cdot\cdot\cdot, p'''_k(\mu'_{k})\rangle.
\end{split}
\end{equation}

The conservation of the probability requires that
\begin{equation}\label{prob}
\langle0|U^\dagger_I(\infty,-\infty)U_I(\infty,-\infty)|0\rangle=1.
\end{equation}
With the expression (\ref{optovac}) we get, the left hand side of this equation could be written as
\begin{equation}
\begin{split}
&\langle0|U^\dagger_I(\infty,-\infty)U_I(\infty,-\infty)|0\rangle \\
&=\sum_{n=0}^\infty\sum_{m=0}^\infty\sum_{l=0}^\infty\sum_{k=0}^\infty e^\alpha
 \mathcal{N}_n^{(0)} \mathcal{N}_m^{(1)} \mathcal{N}_l^{(+)} \mathcal{N}_k^{(-)},
\end{split}
\end{equation}
where
\begin{eqnarray}
  \mathcal{N}_n^{(0)} \!\!\!&=&\!\! \!
      \frac{\tilde{g}^{2n}}{(n!)^2} \int\frac{dp_1^3 |\tilde{\rho}(p_1,\mu_{(0)})|^2}{(2\pi)^{3}2\omega_{\vec{p}_1,\mu_{(0)}}}\cdot\cdot\cdot\frac{dp_n^3 |\tilde{\rho}(p_n,\mu_{(0)})|^2}{(2\pi)^{3}2\omega_{\vec{p}_n,\mu_{(0)}}}n!, \nonumber \\
  \mathcal{N}_m^{(1)} \!\!\!&=&\!\!  \!
     \frac{\tilde{g}^{2m}}{(m!)^2} \!\!\int \!\! \frac{dp'^3_1 |\tilde{\rho}(p'_1,\mu_{(1)})|^2}{(2\pi)^{3}2\omega_{\vec{p'}_1,\mu_{(1)}}}\cdot\cdot\cdot\frac{dp'^3_m |\tilde{\rho}(p'_m,\mu_{(1)})|^2}{(2\pi)^{3}2\omega_{\vec{p'}_m,\mu_{(1)}}}m!, \nonumber \\
  \mathcal{N}_l^{(+)} \!\!\!&=&\!\!\! \frac{\tilde{g}^{2l}}{(l!)^2} \int\frac{dp''^3_1 d\mu_1 |\tilde{\rho}_+(p''_1,\mu_1)|^2}{(2\pi)^{3}2\omega_{\vec{p''}_1,\mu_1}}\times\cdot\cdot\cdot~ \nonumber\\
     &\times &\!\!\! \frac{dp''^3_l d\mu_l |\tilde{\rho}_+(p''_l,\mu_l)|^2}{(2\pi)^{3}2\omega_{\vec{p''}_l,\mu_l}}l!,  \nonumber \\
  \mathcal{N}_k^{(-)} \!\!\!&=&\!\!\!  \frac{\tilde{g}^{2k}}{(k!)^2} \int\frac{dp'''^3_1 d\mu'_1 |\tilde{\rho}_-(p'''_1,\mu'_1)|^2}{(2\pi)^{3}2\omega_{\vec{p'''}_1,\mu'_1}}\times\cdot\cdot\cdot~ \nonumber\\
    &\times&\!\!\! \frac{dp'''^3_k d\mu'_k
    |\tilde{\rho}_-(p'''_k,\mu'_k)|^2}{(2\pi)^{3}2\omega_{\vec{p'''}_k,\mu'_k}}k! .
\end{eqnarray}
One can then get $\alpha$ from Eq. (\ref{prob})
\begin{equation}
\begin{split}
\alpha&=\int\frac{dp_1^3\tilde{g}^2 |\tilde{\rho}(p_1,\mu_{(0)})|^2}{(2\pi)^{3}2\omega_{\vec{p}_1,\mu_{(0)}}}+\int\frac{dp'^3_1 \tilde{g}^2 |\tilde{\rho}(p'_1,\mu_{(1)})|^2}{(2\pi)^{3}2\omega_{\vec{p'}_1,\mu_{(1)}}}\\
&+\int\frac{dp'^3_1 \tilde{g}^2 |\tilde{\rho}(p'_1,\mu_{(1)})|^2}{(2\pi)^{3}2\omega_{\vec{p'}_1,\mu_{(1)}}}\\
&+\int\frac{dp'''^3_1 d\mu'_1 \tilde{g}^2 |\tilde{\rho}_-(p'''_1,\mu'_1)|^2}{(2\pi)^{3}2\omega_{\vec{p'''}_1,\mu'_1}}.
\end{split}
\end{equation}

The number of particles corresponding to all states of the quantum fields we will detect is
\begin{equation}
\begin{split}
\langle N \rangle&=e^\alpha\sum_{n=0}^\infty\sum_{m=0}^\infty\sum_{l=0}^\infty\sum_{k=0}^\infty(n+m+l+k)\\
&\times \mathcal{N}_n^{(0)} \mathcal{N}_m^{(1)} \mathcal{N}_l^{(+)} \mathcal{N}_k^{(-)},
\end{split}
\end{equation}
Now we can return to the reality. We assume that the energy scale of our present accelerator is below $\mu_{(1)}$. Thus, at present energy scale only the ground state of the quantum fields could be detected and we find that the world is four-dimensional, according to previous interpretation on the ground state. In the circumstance that only the ground state could be detected, the number of particles is given by
\begin{equation}\label{number0}
\begin{split}
\langle N_{(0)}\rangle
    &= e^\alpha\sum_{n=0}^\infty\sum_{m=0}^\infty\sum_{l=0}^\infty\sum_{k=0}^\infty n
  \mathcal{N}_n^{(0)} \mathcal{N}_m^{(1)} \mathcal{N}_l^{(+)} \mathcal{N}_k^{(-)}\\
&=\int\frac{dp_1^3 |\tilde{\rho}(p_1,\mu_{(0)})|^2\tilde{g}^2}{(2\pi)^{3}2\omega_{\vec{p}_1,\mu_{(0)}}}.
\end{split}
\end{equation}
As the energy of the accelerator gets higher and exceeds $\mu_{(2)}$, we'll have the chance to meet the excited state modes of the quantum field. Here we focus on a simple case where we are able to detect the first excited state of the fields. Then the particle number we detect will contain the ground state and first excited state modes. The number of these particles is
\begin{equation}
\begin{split}
&\langle N_{(0)+(1)}\rangle \\
&=e^\alpha\sum_{n=0}^\infty\sum_{m=0}^\infty\sum_{l=0}^\infty\sum_{k=0}^\infty (n+m)
  \mathcal{N}_n^{(0)} \mathcal{N}_m^{(1)} \mathcal{N}_l^{(+)} \mathcal{N}_k^{(-)}\\
&=\int\frac{dp_1^3 |\tilde{\rho}(p_1,\mu_{(0)})|^2\tilde{g}^2}{(2\pi)^{3}2\omega_{\vec{p}_1,\mu_{(0)}}}+\int\frac{dp'^3_1 |\tilde{\rho}(p'_1,\mu_{(1)})|^2\tilde{g}^2}{(2\pi)^{3}2\omega_{\vec{p'}_1,\mu_{(1)}}}.
\end{split}
\end{equation}
To make the result more explicit, we can define the particle number density as the particles generated per volume in the energy space
\begin{equation}
\langle N\rangle=\int d\omega\mathcal{N}(\omega).
\end{equation}
For particles of the ground state, Eq.~(\ref{number0}) could be written as
\begin{equation}
\begin{split}
 \langle N_{(0)}\rangle
&=\int d\omega_{\vec{p}_1,\mu_{(0)}}\frac{\omega_{\vec{p}_1,\mu_{(0)}}|\tilde{\rho}(p_1,\mu_{(0)})|^2\tilde{g}^2}{4\pi^2}.
\end{split}
\end{equation}
Similarly, one could get
\begin{equation}
\begin{split}
\langle N_{(0)+(1)}\rangle
&=\int d\omega_{\vec{p}_1,\mu_{(0)}}\frac{\omega_{\vec{p}_1,\mu_{(0)}}|\tilde{\rho}(p_1,\mu_{(0)})|^2\tilde{g}^2}{4\pi^2}\\
&+\int d\omega_{\vec{p}'_1,\mu_{(1)}}\frac{\sqrt{\omega^2_{\vec{p}'_1,\mu_{(1)}}-\mu_{(1)}^2}|\tilde{\rho}(p'_1,\mu_{(1)})|^2\tilde{g}^2}{4\pi^2}.
\end{split}
\end{equation}
Thus we can read off
\begin{equation}
\mathcal{N}_{(0)}(\omega)=\frac{\omega|\tilde{\rho}(p_1,\mu_{(0)})|^2\tilde{g}^2}{4\pi^2}
\end{equation}
and
\begin{equation}
\begin{split}
\mathcal{N}_{(0)+(1)}(\omega)&=\frac{\omega|\tilde{\rho}(p_1,\mu_{(0)})|^2\tilde{g}^2}{4\pi^2}\\
&+\frac{\sqrt{\omega^2-\mu_{(1)}^2}|\tilde{\rho}(p'_1,\mu_{(1)})|^2\tilde{g}^2}{4\pi^2}.
\end{split}
\end{equation}
Considering that the first excited state has four-dimensional mass $\sqrt{2}a$, we could see that at low energy, the particle number density is given as $\mathcal{N}_{(0)}$ and as the energy exceeds $\sqrt{2}a$, the particle number density would be the form of $\mathcal{N}_{(0)+(1)}$. What's more, one can speculate that with the energy getting higher and higher, it's possible to detect more excited states and there will be more terms in the expression of the particle number density.


Next, we will illustrate the above results by taking the source $\rho$ to be the form $\rho(x)=\delta^{(5)}(x-x_0)$ which represents a classical point source located at $x_0$ that could generates mesons. In reality, this type of interaction usually could be realized by crashing two protons together \cite{Coleman:2011xi} or by giving a sudden kick to a classical electron \cite{Peskin:1995ev}. Substituting this form of $\rho(x)$ into Eq.~(\ref{pho0}), we can get $\tilde{\rho}(p_1,\mu_{(0)})$:
\begin{equation}
\begin{split}
\tilde{\rho}(p_1,\mu_{(0)}) 
&=\sqrt{\frac{2a b^5}{\pi }}\cos^4\frac{ay_0}{b}e^{ip_1x_0}.
\end{split}
\end{equation}
Similarly, we get
\begin{equation}
\begin{split}
\tilde{\rho}(p_1,\mu_{(1)})
=\sqrt{\frac{2a b^5}{\pi }}\cos^4 \frac{a y_0}{b}\tan\frac{a y_0}{b}e^{i p_1 x_0}.
\end{split}
\end{equation}
Then the particle number density could be explicitly given as
\begin{equation}
\mathcal{N}_{(0)}(\omega)=\frac{\omega\tilde{g}^2a b^5}{2\pi^3}\cos^8\frac{ay_0}{b}
\end{equation}
and
\begin{equation}
\begin{split}
\mathcal{N}_{(0)+(1)}(\omega)&=\frac{\omega\tilde{g}^2a b^5}{2\pi^3}\cos^8\frac{ay_0}{b}\\
&+\frac{\sqrt{\omega^2-2 a^2}\tilde{g}^2a b^5}{2\pi^3}\cos^8\frac{ay_0}{b}\tan^2\frac{a y_0}{b}.
\end{split}
\end{equation}
These results indicate that at low energy, the particle number density is proportional energy $\omega$. As energy gets higher and exceeds $\mu_{(1)}=\sqrt{2}a$, the first excited state mode could be detected and the particle number density contains two parts. One of them proportional to $\omega$ is the same as that in the low energy case. The other one is proportional to $\sqrt{\omega^2-2a}$. This term comes from the first excited state and could be regarded as an effect of extra dimension. Note that in this paper, we are unable to decide $a$ from the model itself. In other words, we can't decide at which energy scale the excited state and its effect would emerge. The specific value of $a$ needs to be determined by assumption or other braneworld scenarios. Our calculation predicts that if the energy scale is high enough, what different effect the extra dimension, or say excited states in our model, would give.

\section{summary}\label{secIV}
In this paper, to consider the interaction between matter fields in the thick brane model at the quantum level, we adopted the method of quantum field theory in curved spacetime to quantize scalar matter fields in the background spacetime which could generate  P\"{o}schl-Teller potential. Next, we gave the propagator and considered the self-interaction of scalar fields of the form $\mathcal{H}_I=\sqrt{-g}\tilde{g}\rho(x)\phi(x)$. We used the perturbation procedure to solve this interaction and got the analytic form of the corresponding $S$ matrix. We were subsequently able to get the expectation values of particle numbers of the ground state and first excited state, respectively. We then defined the particle number density and from previous results, we showed that when the energy exceeds $\mu_{(1)}$ the first excited state could be detected and the particle number density would change and contain the effect of the first excited state. Finally we constrained our consideration to the special case in which $\rho(x)=\delta^{(5)}(x-x_0)$. In this case, we got the specific forms of densities of particle numbers. And the results tell us how the first excited state, or say extra dimension, affects the particle number density.

While only scalar fields were considered in this paper, the case of fermion fields and more realistic model of interaction between fermions and photons in the thick brane model will be investigated in our later work.

\section{ACKNOWLEDGMENTS}
This work was supported by the National Natural Science Foundation of China (Grants No. 11875151 and No. 11522541), and the Fundamental Research Funds for the Central Universities (Grants No. lzujbky-2018-k11).

\end{document}